\documentstyle[prb,aps,12pt]{revtex}
\begin{document}
\title{Polaron and bipolaron formation in the Hubbard-Holstein model: 
role of next-nearest neighbor electron hopping}
\author{G.~De~Filippis, V.~Cataudella, G.~Iadonisi, 
V.~Marigliano~Ramaglia, C.~A.~Perroni and F.~Ventriglia}
\address{INFM, Unit\`a di Napoli, Dipartimento di Scienze Fisiche, \\
Universit\`a di Napoli I-80126 Napoli, Italy}
\date{\today}
\maketitle

\begin{abstract}

The influence of  
next-nearest neighbor electron hopping, $t^{\prime}$, on the polaron         
and bipolaron 
formation in a square Hubbard-Holstein model is investigated within a 
variational approach.
The results for electron-phonon and electron-electron correlation functions 
show that a negative value of $t^{\prime}$ induces a strong anisotropy 
in the lattice distortions favoring the formation of nearest neighbor 
intersite bipolaron. The role of $t^{\prime}$, electron-phonon and 
electron-electron interactions is briefly discussed in view of  
the formation of charged striped domains.
\end{abstract}


\newpage 

\section {Introduction}
 
In recent years the experimental evidence in favor of polaronic 
carriers in doped cuprates and in manganese oxide perovskites has grown. 
In manganites a large amount of experimental results, 
ranging from  EXAFS\cite{1} measurements of lattice distortions to 
giant isotope shift of the Curie temperature\cite{2} and to frequency shifts 
of the internal phonon modes,\cite{3} have pointed out the relevance of the 
Jahn-Teller polaron formation\cite{4} beside 
the double and super-exchange magnetic effects.\cite{5} 
Also in cuprates there is strong experimental evidence 
supporting a relevant role of the interaction between charge carriers and  
lattice distortions in addition to the strong electron correlations. 
Optical experiments in the mid-infrared frequency region,\cite{6} atomic pair 
distribution function analysis of the neutron powder diffraction data\cite{7} 
and of the EXAFS signals\cite{8} due to the $Cu-O$ bond distances have shown 
polaronic effects in doped cuprates pointing out a strong 
response of the local structure to the charge state.

This large amount of experimental data has renewed the interest in studying 
problems of electrons interacting with the lattice degrees of freedom. 
In literature several models have been introduced to treat the 
electron-electron (el-el)  and 
electron-phonon (e-ph) interactions in these compounds.
\cite{becca,aligia} 
In this paper we will restrict our attention to one of  
the most simple and frequently considered models for 
the polaron and bipolaron formation: the Holstein-Hubbard model.
The problem of a single tight binding electron coupled to an 
optical local phonon mode has been analyzed in several ways:  
Monte Carlo simulations,\cite{9} 
numerical exact diagonalization of small clusters,\cite{10} 
dynamical mean field theory,\cite{11} density matrix renormalization 
group\cite{12} and variational approaches.\cite{13,romero,var}  
As a common result the ground state energy and the effective 
mass in the Holstein model are continuous functions of the e-ph 
coupling constant and this 
one-body system does not have phase transition.\cite{14} 
Depending on the adiabatic parameter  
the ground state properties can change more or less 
significantly but without breaking the translational symmetry. 
Recently the influence of the Hubbard repulsion on bipolaron formation 
has been investigated\cite{bonca,16} by variational and exact 
diagonalization methods in one and two dimensions. In the adiabatic 
regime the possibility of formation of intersite 
bipolarons has been suggested and it has been shown that 
their mass is significantly reduced 
with respect to the on-site bipolaron.    
 
On the other hand accurate investigations on high temperature 
superconducting materials have shown that some properties of cuprates 
compounds, as the shape of the Fermi surface or the band structure, 
can be explained introducing a next-nearest neighbor (nnn) 
electron hopping term.\cite{new} 
This term is essential for reproducing the experimentally observed behavior  
of the electron band near the $M$ points of the Brillouin 
zone and it allows to handle the differences between electron and hole 
doped materials.
The inclusion 
of nnn electron hopping is expected to affect significantly the behavior 
of the system. For instance, it has been shown that $t^{\prime}$ 
may deeply modify the properties of 
the $t-j$ and Hubbard models and 
that the renormalization of the bare parameters, $t$ and $t^{\prime}$, 
can be very strong also for moderate 
values of the electron correlations.\cite{19}   

In this paper we investigate, within a variational approach, the influence of 
nnn transfer integral on the polaron and bipolaron 
formation in the two dimensional Hubbard-Holstein model. In particular for the 
single electron we use a recently proposed variational approach\cite{13} 
based 
on a linear superposition of Bloch states that describe large and small 
polaron wave functions. 
This approach provides a very good description of the polaron features in any 
regime of the parameters of the Holstein model and  
it does not involve a truncation of the boson
Hilbert space as required by all the numerical techniques. 
The computational effort is very little involving
few variational parameters. 
A variational approach is used also to study the ground state 
bipolaron features. In this case we use a single wave function able to 
interpolate between large and small bipolaron regimes. 
In order to check the validity of the proposed wave 
function we compare our results for the one-dimensional Hubbard-Holstein 
model with those recently published by Bonca 
et al.,\cite{bonca} which are the most accurate bipolaron available 
calculations.  In the 
most interesting regime, characterized by electron and phonon energy scales 
not well separated, the difference between the bipolaron ground state 
energy estimations of the two methods is about 0.1\%: this makes us confident 
of the accuracy of the proposed approach. Next we apply the variational method 
to the two dimensional extended Holstein-Hubbard model.  

In this paper 
we show that a negative value of $t^{\prime}$ induces a relevant  
anisotropy in the lattice displacements associated to the 
polaron formation. A decreasing of $t^{\prime}$, in the range of 
parameters of physical interest, strengthens the e-ph correlations 
along the $(1,0)$ and $(0,1)$ directions and reduces the lattice displacements 
along the (1,1) and (1,-1) directions. 
Moreover a negative value of $t^{\prime}$ favors 
the mobile intersite bipolaron formation along the $(1,0)$ and $(0,1)$ 
directions and, when this does not happen, the 
el-el effective potential shows 
a strong dependence on the spatial 
directions. Although these results concern the case of only two electrons 
interacting with local phonon modes, we believe that the formation of 
intersite bipolaron in the (1,0) direction suggests the possibility that, 
at physical relevant doping in the manganites and cuprates, the e-ph 
interaction together with anisotropy can favor the formation 
of charged stripes. Of course, further investigation is needed to support 
this idea in specific materials.    

The paper is organized in the following way. In section 2 the model is 
introduced. In section 3 
the variational approach for a single polaron is discussed. 
In section 4 the bipolaron properties within 
the one-dimensional Holstein-Hubbard model are analyzed by means of a 
variational method and are successful compared with the best available results 
recently published by Bonca et al..\cite{bonca} Successively the variational 
approach is extended to study the bipolaron features within 
the two-dimensional Hubbard-Holstein model. In section 5 the numerical results 
are presented and the role of $t^{\prime}$, e-ph and 
el-el interactions is briefly discussed in view of  
the formation of charged striped domains.  

\section {The model}

The two-dimensional extended Hubbard-Holstein model is described by the 
Hamiltonian: 
\begin{eqnarray}
H=&& -t\sum_{i,\delta,\sigma} c^{\dagger}_{i+\delta,\sigma}c_{i,\sigma}
-t^{\prime}\sum_{i,\delta^{\prime},\sigma}
c^{\dagger}_{i+\delta^{\prime},\sigma}c_{i,\sigma} 
+\omega_0\sum_{\vec{q}
}a^{\dagger}_{\vec{q}}a_{\vec{q}}  \nonumber \\
&&+\frac{g \omega_0}{\sqrt{N}} \sum_{i,\sigma,\vec{q}}
c^{\dagger}_{i,\sigma}c_{i,\sigma} \left(a_{\vec{q}} e^{i \vec{q} \cdot 
\vec{R}_i}
+a^{\dagger}_{\vec{q}} e^{-i \vec{q} \cdot \vec{R}_i}\right)+
U\sum_{i}n_{i,\uparrow}n_{i,\downarrow}~.  
\label{1r}
\end{eqnarray}

In the Eq.(\ref{1r}) $c^{\dagger}_{i}$ denotes the electron creator 
operator at site $i$, whose position vector is indicated by 
$\vec{R}_{i}$, $a^{\dagger}_{\vec{q}}$ represents the creation operator 
for phonon with wave number $\vec{q}$, $t$ and $t^{\prime}$ are, 
respectively, the transfer integral between 
nearest and next nearest neighbor sites, $\vec{\delta}$ and 
$\vec{\delta}^{\prime}$ indicate, respectively, 
the next and next-nearerst neighbors, 
$\omega_0$ is the frequency of the optical local 
phonon mode, $U$ represents the Hubbard interaction for electrons on the same 
site and $M_{q}$ indicates the e-ph matrix 
element. In the Holstein model (short range e-ph interaction) 
$M_{q}$ assumes the form: 
\begin{equation}
M_{q}=\frac{g}{\sqrt{N}} \omega_0 .
\label{106r}
\end{equation}
Here $N$ is the number of 
lattice sites. 

\section {Polaron variational approach} 

In previous papers,\cite{13} we have developed a variational approach 
based on a linear superposition of Bloch states that represent the large and 
small polaron wave functions. 
These two wave functions are chosen as 
translationally invariant Bloch states, they are 
obtained from 
localized states centered on different lattice sites,  
just like a band state is related to atomic orbitals, and 
they provide a very accurate description of 
the two asymptotic regimes of weak and strong e-ph coupling: 
\begin{equation}
|\psi^{(\alpha)}_{\vec{k}}\rangle=
\frac{1}{\sqrt{N}}\sum_{n}e^{i\vec{k}\cdot 
\vec{R}_n}|\psi^{(\alpha)}_{\vec{k}}(\vec{R}_n)\rangle
\label{12r}
\end{equation}
where 
\begin{equation}
|\psi^{(\alpha)}_{\vec{k}}(\vec{R}_n)\rangle
=e^{\sum_{\vec{q}}\left[f^{(\alpha)}_{\vec{q}}(\vec{k})a_{\vec{q}}
e^{i\vec{q}\cdot \vec{R}_n}
-h.c.\right]}
\sum_{m}
\phi^{(\alpha)}_{\vec{k}}(\vec{R}_m)
c^{\dagger}_{n+m}
\left(|0\rangle_{el} \otimes  |0\rangle_{ph}\right)~.
\label{13r}
\end{equation} 

In the Eqs.(\ref{12r},\ref{13r}) 
the apex $\alpha$ indicates the large ($\alpha=l$) and small ($\alpha=s$) 
polaron wave 
function, 
$|0\rangle_{el}$ and $|0\rangle_{ph}$ denote the 
electron and boson vacuum states and 
$\phi^{(\alpha)}_{\vec{k}}(\vec{R}_m)$ are variational 
parameters such that   
$\sum_{m}|\phi^{(\alpha)}_{\vec{k}}(\vec{R}_m)|^2=1~$. 
These two wave functions are characterized by different phonon 
distribution functions: 
\begin{equation}
f^{(l)}_{\vec{q}}(\vec{k})=
\frac{g\omega_0 / \sqrt{N}}{\omega_0+E_b(\vec{k}+\vec{q})-E_b(\vec{k})}
\label{225r}
\end{equation}
and 
\begin{equation}
f^{(s)}_{\vec{q}}(\vec{k})=
\frac{g}{\sqrt{N}}\sum_{i}e^{i\vec{q}\cdot \vec{R}_i}
|\phi^{(s)}_{\vec{k}}(\vec{R}_i)|^2.
\label{22r}
\end{equation}
Here $E_b(\vec{q})$ is the free electron band energy and 
the variational parameters 
$\phi^{(\alpha)}_{\vec{k}}(\vec{R}_m)$
take into account  
the broadening of the electron wave function. In this paper we restrict the 
sum in Eq.(\ref{13r})    
till to fifth neighbors. It is evident that 
the large polaron wave function takes into account 
the average effect of the correlation introduced by the electron 
recoil (Eq.\ref{225r}), effect absent in the small polaron phonon 
distribution function. 

We have shown\cite{13}   
that these wave functions, 
far away from 
the two asymptotic regimes, are not orthogonal and the  
off-diagonal matrix elements of the Holstein Hamiltonian are not zero. 
This suggests that the lowest 
state of the system is made of a mixture of the large and small polaron 
solutions and so justifies  
the idea to use a variational method to determine the ground state energy  
by considering as trial state 
a linear superposition of the wave functions describing the two types of 
previously discussed polarons. At $t^{\prime}=0$ the comparison of the 
numerical results with the data of the Density Matrix Renormalization 
group\cite{12} and Global Local Variational\cite{romero} methods has shown 
the great accuracy of the proposed approach.   

\section {Bipolaron variational approach}

\subsection {One-dimensional case}

Regarding the bipolaron, the most general wave function of two electrons 
in a periodic potential interacting with the longitudinal optical phonons 
and with each other through the Coulomb force can be written as: 
\begin{equation}
|\psi_{\vec{k}}\rangle=\frac{1}{N}\sum_{n_1,n_2}
e^{i\vec{k} \cdot \frac{\left(\vec{R}_{n_1}+\vec{R}_{n_2}\right)}{2}}
|\psi_{\vec{k}}(\vec{R}_{n_1},\vec{R}_{n_2})
\rangle~. 
\label{100r}
\end{equation}
$|\psi_{\vec{k}}\rangle $ is a state that is multiplied 
by the factor $e^{i\vec{k} \cdot \vec{R}_m}$
under a lattice vector $\vec{R}_m$ translation 
and $\vec{k}$ is the  
Bloch state wave number.   
The Eq.(\ref{13r}), i.e. the polaron wave function component describing the 
charge carrier distributed around the site $\vec{R}_n$, is the key ingredient 
to build the bipolaron wave function component 
$|\psi_{\vec{k}}(\vec{R}_{n_1},\vec{R}_{n_2})\rangle$. Fixed the 
relative distance between the centers of the two electrons 
($\vec{R}_{n_1}-\vec{R}_{n_2}$), we adopt a trial wave function for the 
singlet state that is proportional to the product of two polaron 
wave functions centered on $\vec{R}_{n_1}$ and $\vec{R}_{n_2}$ sites:  
\begin{eqnarray}
|\psi_{\vec{k}}(\vec{R}_{n_1},\vec{R}_{n_2})\rangle&&= 
\gamma_{\vec{k}}(\vec{R}_{n_1}-\vec{R}_{n_2})
e^{
{\sum_{\vec{q}}}
\{ 
h_{\vec{q}}(\vec{k},\vec{R}_{n_1}-\vec{R}_{n_2})
a_{\vec{q}}
\left[
e^{i\vec{q}\cdot \vec{R}_{n_1}}
+e^{i\vec{q}\cdot \vec{R}_{n_2}}
\right]
-h.c.
\}
}\nonumber\\
&&\sum_{m_1}
\phi_{\vec{k}}(\vec{R}_{m_1},\vec{R}_{n_1}-\vec{R}_{n_2}) 
c^{\dagger}_{n_1+m_1,\uparrow}
\sum_{m_2}
\phi_{\vec{k}}(\vec{R}_{m_2},\vec{R}_{n_1}-\vec{R}_{n_2})
c^{\dagger}_{n_2+m_2,\downarrow}
\left(|0\rangle_{el} \otimes |0\rangle_{ph}\right)
\nonumber
\end{eqnarray}
where $\phi$,   
$\gamma$ and $h$ are variational 
functions. In particular, 
$\gamma_{\vec{k}}(\vec{R}_{n_1}-\vec{R}_{n_2})$ gives the weight 
of the bipolaron wave function component 
with the centers of the two charge carriers 
at distance $|\vec{R}_{n_1}-\vec{R}_{n_2}|$. 
For the phonon distribution function $h$ we assume the 
following form:
\begin{eqnarray}
h_{\vec{q}}(\vec{k},\vec{R}_{n_1}-\vec{R}_{n_2})=
\frac{g\omega_0 d(\vec{k},\vec{R}_{n_1}-\vec{R}_{n_2})/\sqrt{N}}
{\omega_0+h_1(\vec{k}) \left( E_b(t,\vec{k}+\vec{q})-E_b(t,\vec{k})
\right)} \nonumber
\end{eqnarray}
that is able to interpolate between the asymptotic expressions 
of large ($h_1=d=1$) and small 
($h_1=0$ and $d=1$) polaron. Here 
$E_b(t,\vec{q})=-2t\cos(q_xa)$ is the free electron band energy,  
$a$ is the lattice constant  
and $d(\vec{k},\vec{R}_{n_1}-\vec{R}_{n_2})$ and  
$h_1(\vec{k})$ are variational functions. 
We note that we have not used a linear superposition of 
polaron wave functions to build the bipolaron wave function, but, following 
Toyozawa,\cite{toyo} a single wave function able to interpolate between 
the two asymptotic regimes, as shown by Romero et al..\cite{romero}  This 
approach allows us to reduce the number of variational parameters, it 
requires, to be implemented, a very little computational effort and it 
provides a very good description of the bipolaron ground state properties as 
will be shown in the following.

The coefficient 
$\phi$ has been chosen such that it
takes into account the broadening of the electron wave function till 
to third neighbors. Therefore,  
for any fixed relative distance between the centers of the 
two charge carriers, we introduce three independent variational parameters: 
\[ \phi_{\vec{k}}(\vec{R}_m,\vec{R}_{n_1}-\vec{R}_{n_2})= 
\left\{ 
\begin{array}{ll} 
\alpha(|\vec{R}_{n_1}-\vec{R}_{n_2}|) & \mbox{if $|\vec{R}_m|=0$} \\
\beta(|\vec{R}_{n_1}-\vec{R}_{n_2}|) & \mbox{if $|\vec{R}_m|=a$} \\
\gamma(|\vec{R}_{n_1}-\vec{R}_{n_2}|) & \mbox{if $|\vec{R}_m|=2a$} \\
\delta(|\vec{R}_{n_1}-\vec{R}_{n_2}|) & \mbox{if $|\vec{R}_m|=3a$} \\
0 & \mbox{otherwise}
\end{array} 
\right. 
\]
with $\alpha^2+2 (\beta^2+\gamma^2+\delta^2)=1$. 
In this paper, to simplify the numerical calculations, we have chosen 
$\alpha$, $\beta$, $\gamma$, $\delta$ independent on 
$|\vec{R}_{n_1}-\vec{R}_{n_2}|$ for $|\vec{R}_{n_1}-\vec{R}_{n_2}|\ge 3a$. 
Likewise the function 
$d(\vec{k},\vec{R}_{n_1}-\vec{R}_{n_2})$ assumes the following form: 
\[ d(\vec{k},\vec{R}_{n_1}-\vec{R}_{n_2})=
\left\{ 
\begin{array}{ll} 
d_0(\vec{k}) & \mbox{if $|\vec{R}_{n_1}-\vec{R}_{n_2}|=0$} \\
d_1(\vec{k}) & \mbox{if $|\vec{R}_{n_1}-\vec{R}_{n_2}|=a$} \\
d_2(\vec{k}) & \mbox{if $|\vec{R}_{n_1}-\vec{R}_{n_2}|=2a$} \\
d_3(\vec{k}) & \mbox{otherwise} 
\end{array} 
\right. 
\]
where $d_0(\vec{k})$, $d_1(\vec{k})$, $d_2(\vec{k})$ and $d_3(\vec{k})$ 
are variational parameters. Finally we choose the following form for the 
function $\gamma_{\vec{k}}(\vec{R}_{n_1}-\vec{R}_{n_2})$
\[ \gamma_{\vec{k}}(\vec{R}_{n_1}-\vec{R}_{n_2})=
\left\{ 
\begin{array}{ll} 
\gamma_0(\vec{k}) & \mbox{if $|\vec{R}_{n_1}-\vec{R}_{n_2}|=0$} \\
\gamma_1(\vec{k}) & \mbox{if $|\vec{R}_{n_1}-\vec{R}_{n_2}|=a$} \\
\gamma_2(\vec{k}) & \mbox{if $|\vec{R}_{n_1}-\vec{R}_{n_2}|=2a$} \\
\gamma_3(\vec{k})e^{-\gamma_4(\vec{k})|\vec{R}_{n_1}-\vec{R}_{n_2}|} 
& \mbox{otherwise} 
\end{array} 
\right. 
\]
where $\gamma_0(\vec{k})$, $\gamma_1(\vec{k})$, $\gamma_2(\vec{k})$ and 
$\gamma_3(\vec{k})$ are variational parameters.  
The minimization of the quantity 
$\frac{\langle\psi_{\vec{k}}|H|\psi_{\vec{k}}\rangle}
{\langle\psi_{\vec{k}}|\psi_{\vec{k}}\rangle}$  
has been performed by making use of a routine 
based on a standard Newton algorithm. 

In Fig.1 we plot the energy difference between the two-particle 
ground state and twice the one-particle ground state as function of the 
Hubbard repulsion $U$ at $t=\omega_0$ and $g=1$. The comparison with 
the numerical results recently published by Bonca et al.\cite{bonca} shows 
the accuracy of the proposed variational approach. For example, the present 
method's estimate of the bipolaron ground state energy for $g=1$, $t=\omega_0$ 
and $U=0$ in the thermodynamic limit is $E_0=-5.4185$ that is 
in excellent agreement with the Bonca's 
result: $E_0=-5.4246$, the difference being about 0.1\%. In Fig.1b and Fig.1c 
the electron-electron correlation function $\rho(\vec{R}_m)=\langle 
\psi_{\vec{k}=0}|
\sum_{i}n_{i,\uparrow}n_{i+m,\downarrow} |\psi_{\vec{k}=0} \rangle/
\langle\psi_{\vec{k}=0}|\psi_{\vec{k}=0}\rangle$ 
is plotted for two different values of $U$ at 
$t=\omega_0$ and $g=1$. The results are compared with the data reported 
by Bonca et al.\cite{bonca}. The comparison shows that the used wave function 
provides not only an excellent estimation of the ground state energy 
but gives a high accurate description of the bipolaron ground state 
properties. In the next subsection we 
extend the proposed approach to the two dimensional case. 

\subsection {Two-dimensional case}

Here we extend the variational approach to explore the bipolaron features
for electrons interacting with longitudinal 
optical phonons and with each other 
in a two-dimensional lattice.
As in the one-dimensional case, 
we adopt a trial wave function for the singlet state which is   
a boson coherent state multiplied by the product 
of linear superpositions of Wannier wave functions: 
\begin{eqnarray}
|\psi_{\vec{k}}(\vec{R}_{n_1},\vec{R}_{n_2})\rangle&&= 
\gamma_{\vec{k}}(\vec{R}_{n_1}-\vec{R}_{n_2})
e^{
{\sum_{\vec{q}}}
\{ 
h_{\vec{q}}(\vec{k},\vec{R}_{n_1}-\vec{R}_{n_2})
a_{\vec{q}}
\left[
e^{i\vec{q}\cdot \vec{R}_{n_1}}
+e^{i\vec{q}\cdot \vec{R}_{n_2}}
\right]
-h.c.
\}
}\nonumber\\
&&\sum_{m_1}
\phi_{\vec{k}}(\vec{R}_{m_1},\vec{R}_{n_1}-\vec{R}_{n_2}) 
c^{\dagger}_{n_1+m_1,\uparrow}
\sum_{m_2}
\chi_{\vec{k}}(\vec{R}_{m_2},\vec{R}_{n_1}-\vec{R}_{n_2})
c^{\dagger}_{n_2+m_2,\downarrow}
\left(|0\rangle_{el} \otimes |0\rangle_{ph}\right)
\nonumber
\end{eqnarray}
where $\phi$, $\chi$,  
$\gamma$ and $h$ are variational 
functions, with 
$\chi_{\vec{k}}(\vec{R}_{m},\vec{R}_{n_1}-\vec{R}_{n_2})=
\phi_{\vec{k}}(\vec{R}_{m},\vec{R}_{n_2}-\vec{R}_{n_1})$. For 
the phonon distribution function we assume the 
following form:
\begin{eqnarray}
h_{\vec{q}}(\vec{k},\vec{R}_{n_1}-\vec{R}_{n_2})=
\frac{g\omega_0 d(\vec{k},\vec{R}_{n_1}-\vec{R}_{n_2})/\sqrt{N}}
{\omega_0+h_1(\vec{k}) \left( E_b(t,\vec{k}+\vec{q})-E_b(t,\vec{k})
\right) 
+h_2(\vec{k}) \left( E_b(t^{\prime},\vec{k}+\vec{q})- E_b(t^{\prime},\vec{k})
\right)}\nonumber
\end{eqnarray}
that is able to interpolate between the asymptotic expressions 
of large ($h_1=h_2=d=1$) and small  
($h_1=h_2=0$ and $d=1$) polaron. Here 
$E_b(t,\vec{q})=-2t \left(\cos(q_xa)+\cos(q_ya)\right)$,
$E_b(t^{\prime},\vec{q})=-4t^{\prime}\cos(q_xa)\cos(q_ya)$ 
and $d(\vec{k},\vec{R}_{n_1}-\vec{R}_{n_2})$, 
$h_1(\vec{k})$ and $h_2(\vec{k})$ are variational functions. The coefficients 
$\phi$ and $\chi$ have been chosen such that they 
take into account the broadening of the electron wave function till 
to next-nearest neighbors, i.e. the sums in the expression of 
$|\psi_{\vec{k}}(\vec{R}_{n_1},\vec{R}_{n_2})\rangle$ are restricted 
till to next nearest neighbors.  
Furthermore $\phi$ and $\chi$ have the bipolaron 
symmetry for   
relative distances $|\vec{R}_{n_1}-\vec{R}_{n_2}| \le 2\sqrt{2} a$ 
(fifth neighbors), while for $|\vec{R}_{n_1}-\vec{R}_{n_2}| > 2\sqrt{2} a$ 
we assume, in analogous way to the one-dimensional case, the following form 
for $\phi$: 
\[ \phi_{\vec{k}}(\vec{R}_m,\vec{R}_{n_1}-\vec{R}_{n_2})= 
\left\{ 
\begin{array}{ll} 
\alpha & \mbox{if $|\vec{R}_m|=0$} \\
\beta & \mbox{if $\vec{R}_m=\vec{\delta}$} \\
\gamma & \mbox{if $\vec{R}_m=\vec{\delta}^{\prime}$} \\
0 & \mbox{otherwise}
\end{array} 
\right.
\] 
with $\alpha^2+4 (\beta^2+\gamma^2)=1$ and  
$\alpha$, $\beta$, $\gamma$ independent on $(\vec{R}_{n_1}-\vec{R}_{n_2})$. 
The on site, nearest neighbor and next-nearest neighbor values of the 
functions $\gamma_{\vec{k}}(\vec{R}_{n_1}-\vec{R}_{n_2})$ and 
$h_{\vec{q}}(\vec{k},\vec{R}_{n_1}-\vec{R}_{n_2})$  
have been determined variationally whereas for relative  
distances between the centers of the two charge carriers such that 
$|\vec{R}_{n_1}-\vec{R}_{n_2}| > \sqrt{2}a$ 
(next-nearest neighbors) we have used  
asymptotic expressions with parameters fixed by the variational 
approach as in the above-introduced one-dimensional case.  
Here we take into account the polaron and bipolaron ground state
($\vec{k}=0$).

\section {The results}
 
In Fig.2 we plot the polaron ground state energy $E$, 
the mean phonon number $N$,
the spectral weight $Z$ 
and the polaron kinetic energy  
$K$, in units of the bare electron kinetic energy, 
as a function of the e-ph coupling constant, at $t/\omega_0=2$, 
for different values of nnn transfer integral $t^{\prime}$.
At $t^{\prime}=0$ there is a sharp transition between the large and small 
polaron solutions for $\lambda \simeq 1$, where 
$\lambda=g^2 \omega_0/4t$ is the ratio between the small polaron binding 
energy and the energy gain of an itinerant electron on a rigid lattice.
In the weak coupling regime $K$ is $\simeq 1$ and $N$ is $\simeq 0$ 
so that the electron is slightly affected by the interaction with 
the phonons: its bare mass presents weak 
renormalization. In the opposite regime the 
polaron band collapses:  
the average number of phonons increases, the kinetic energy reduces and  
asymptotically tend to the values predicted by the strong coupling 
perturbation theory ($N\rightarrow g^2$, $K\rightarrow e^{-g^{2}}$). 
The ground state spectral weight shows a sharp transition at 
$\lambda\simeq 1$ but 
it is evident the presence of a wide range of values of the e-ph 
coupling constant where $Z$ is significantly smaller than the unity but  
not negligible: here the ground state properties are those 
characteristic of an 
electron weakly affected by the e-ph interaction but 
a large part 
of the single-particle spectral weight lies at higher energies  
(intermediate polaron phase\cite{13}).  
Decreasing the value of $t^{\prime}$ ($t^{\prime}<0$), this intermediate 
regime  
becomes more important with respect to the small and large 
polaron phases: 
the spectral weight is equally distributed between 
ground and all the excited states and the polaron ground state features 
are well described by a linear 
superposition of  
$|\psi^{(l)}_{\vec{k}}\rangle$ and $|\psi^{(s)}_{\vec{k}}\rangle$. 
In the weak coupling regime, 
for a fixed value of the e-ph coupling constant, decreasing $t^{\prime}$   
the mean phonon number grows, the ground state spectral weight and the 
kinetic energy decrease indicating a greater polaron localization.  
The strong coupling regime shows, instead, a non 
monotonous behavior as function of $t^{\prime}$ that can be explained 
in terms of the third order of the strong coupling 
perturbation theory. In fact, at this order  
the small polaron binding  energy is the sum of three 
contributions: the first term is $-g^2 \omega_0$, the second and the third 
ones are proportional respectively to $-(t^2+t^{\prime 2})/g^2\omega_0$ 
and $-t^2 t^{\prime}/g^3\omega_0^2$.   

The effect of nnn transfer integral on the lattice 
displacements associated to the polaron formation, $S(\vec{R}_m)=\langle
\sum_{i}n_i\left(a_{i+m}+a^{\dagger}_{i+m}\right)\rangle$, 
is particularly 
interesting. A negative value of $t^{\prime}$ induces a strong anisotropy 
in the lattice distortions along the (1,1) and (1,0) directions  
as it results from Fig.3. 
In the range of parameters of physical interest, for a fixed value of the 
e-ph interaction, $S(1,0)/S(0,0)$ 
and $S(1,1)/S(0,0)$ decrease in a very different way with $t^{\prime}$.  
In fact,  the ratio 
$S(1,1)/S(1,0)$ reduces in a dramatic way indicating very strong 
correlations  between electron positions and lattice displacements along the 
symmetry axes of the crystal. 

A similar trend is revealed by correlation 
functions of two interacting electrons. 
We have investigated the bipolaron formation in the adiabatic regime where  
the retardation effect of the e-ph 
interaction may favor the rise of more extended electron bound states 
and in particular the mobile intersite bipolaron formation.\cite{bonca,16} 
Fig4.a shows 
the phase diagram for the transition from unbound polarons to 
bipolarons at fixed $t$ and $U$. 
We stress that the presence of a negative 
nnn electron hopping extends the region of g-values where 
the intersite mobile bipolaron formation is favored. 
In Fig4.b the density-density correlation function 
$\rho(\vec{R}_m)=\langle \psi|
\sum_{i}n_{i,\uparrow}n_{i+m,\downarrow} |\psi \rangle/
\langle\psi|\psi\rangle$ is plotted. 
When the intersite bipolaron formation takes place, 
the probability to find the two interacting electrons along the $(1,0)$  
and $(0,1)$ directions 
exceeds the $90\%$ and the maxima of $\rho(\vec{R}_m)$ are  
located at $\vec{R}_m=\{(\pm 1,0),(0,\pm 1)\}$. When the bipolaron does 
not form,  
both along $(1,0)$ and $(1,1)$ directions $\rho(\vec{R}_m)$ does 
not show any structure and  
the typical distance 
between the two particles is of the order of the maximum allowed separation. 
In this case the behavior of the effective potential is enlightening.   
At the lowest order ($\phi(\vec{R}_m)=\delta_{m,0}$),  
$V_{eff}(\vec{R}_m)=U \delta_{m,0}+2 \omega_0 
\sum_{\vec{q}}(h^2_{\vec{q}}(\vec{R}_m)
-2g 
h_{\vec{q}}(\vec{R}_m)/\sqrt{N})(1+\cos{\vec{q}\cdot \vec{R}_m})$. At $t^{\prime}=0$, 
the superposition of the polarization clouds produces an 
isotropic attractive force between the two charge carriers; decreasing 
$t^{\prime}$, the effective potential is more attractive 
along the $(1,0)$ direction and less attractive along $(1,1)$ direction  
due to the anisotropy in the lattice distortions generated by each polaron, 
as it results from Fig4.c. 
   
Finally we discuss the possible consequences of our results for the 
formation of striped charge distributions in the Hubbard-Holstein model. 
First of all, since $t^{\prime}$ favors the formation of nn intersite 
bipolaron, we expect that the nnn electron hopping may support, 
for realistic values of the doping, 
the formation of striped structures 
arranged along characteristic directions of the crystal. 
Furthermore we stress the existence of an intermediate phase  
where small and large polaron properties coexist. This property of the single 
polaron, as discussed in previous papers\cite{26}, 
can support, in a many body problem,
a first order phase transition with coexistence of  
charged striped domains characterized by 
different densities and lattice deformations.
The possibility 
of striped structures, fluctuating and short range ordered, and 
the eventuality of coexistence of two types of charge carriers forming striped 
domains    
have been suggested as possible scenarios for cuprates\cite{25,8,17}. 
There are also 
strong experimental evidences from 
EXAFS,\cite{8,17} neutron scattering\cite{23} and X-ray scattering\cite{24} 
in favor of the existence of charged striped structures. 
In both cases 
the interplay of $t^{\prime}$, el-ph and el-el interactions 
may significantly affect the 
physics of the system. Finally we want to stress that the 
Hamiltonian (\ref{1r}) represents one of the simplest models to take 
into account the el-el and e-ph interactions and that in the case of the 
cuprates one needs to consider more realistic models of 
electron-phonon interaction, for example, models  
where the $Cu-O$ hoppings are modified by the 
ion displacements.\cite{aligia} However the Holstein-Hubbard model is 
expected to provide useful informations on the physics of these systems, 
at least on a qualitatively level, and it 
can be an interesting starting point for studying their caractheristic 
features.      

In conclusion we have investigated within a variational approach the influence 
of next-nearest neighbor transfer integral on the polaron and bipolaron 
formation in the two dimensional Hubbard-Holstein model. It has been shown 
that a negative value of $t^{\prime}$ induces a relevant anisotropy in the 
lattice displacements associated to the polaron formation and favors the 
mobile intersite bipolaron presence along the $(1,0)$ and $(0,1)$ directions. 
Finally we have discussed the relevance of these results for the formation 
of striped structures arranged along characteristic direction of the crystal 
and for the coexistence of charged domains characterized by different 
densities and lattice deformations.

\section*{Figure captions}
\begin{description}
\item  {Fig.1} (a) The bipolaron binding energy is plotted as 
function of $U$ for a 
one-dimensional lattice of $36$ 
sites with periodic boundary conditions. The values 
of the parameters are $t=\omega_0$ and $g=1$. The stars indicate the data 
kindly provided by J. Bonca; (b) and (c): the electron-electron correlation 
function is plotted at $g=1$ and $t=\omega_0$ for two different values of $U$: 
$U=0$ in (b) and $U=1.5$ in (c). The energies are given in units of 
$\omega_0$. The stars represent the data kindly provided by J. Bonca. 
\item  {Fig.2} 
The polaron ground state energy, (a), the mean phonon number, (b),
the spectral weight, (c),  
and the mean value of the hopping term, 
in units of the bare electron kinetic energy, (d), 
are reported, for a two dimensional lattice in the thermodynamic limit 
($N\rightarrow \infty$), 
at $t/\omega_0=2$ for different values of $t^{\prime}$: 
$t^{\prime}/\omega_0=0$ 
(solid line), $t^{\prime}/\omega_0=-0.4$ (dashed line), 
$t^{\prime}/\omega_0=-0.6$
(dotted line), $t^{\prime}/\omega_0=-0.8$ (dashed-dotted line).    
The energies are given in units of $\omega_0$.
\item  {Fig.3} (a): 
$d_{m,0}=S(|\vec{R}_m|)/S(|\vec{R}_m|=0)$ 
($|\vec{R}_m|$ indicates the modulus of the 
vector $\vec{R}_m$ in units of the lattice parameter $a$) is reported at 
$t/\omega_0=2$ 
and $g=2$ for two different values of $t^{\prime}$; 
(b): $d_{1,0}=S(1,1)/S(1,0)$ 
is reported  at $t/\omega_0=2$ for different values of $t^{\prime}$: 
$t^{\prime}/\omega_0=0$ 
(solid line), $t^{\prime}/\omega_0=-0.4$ (dashed line), 
$t^{\prime}/\omega_0=-0.6$
(dotted line), $t^{\prime}/\omega_0=-0.8$ (dashed-dotted line); 
(c): $d_{1,0}=S(1,1)/S(1,0)$ 
is reported  at $t/\omega_0=2$ for different values of $g$: 
$g=2$ 
(solid line), $g=2.5$ (dashed line), 
$g=2.65$
(dotted line), $g=2.75$ (dashed-dotted line). The data refer to two 
dimensional lattice in the thermodynamic limit ($N\rightarrow\infty$).   
\item  {Fig.4} (a): the phase diagram for $t/\omega_0=4$ and 
$U/\omega_0=30$ on a square 
lattice $10$x$10$ with periodic boundary conditions;  
(b): the density-density correlation function $\rho(|\vec{R}_m|)$ is plotted  
at $t/\omega_0=4$, $U/\omega_0=30$ for $g=3.5$ and $t^{\prime}/\omega_0=-1$;
(c): $V_{eff}(|\vec{R}_m|)$, measured by twice the one-particle potential 
energy, is reported 
at $t/\omega_0=4$, $U/\omega_0=30$, $g=2.8$ for two different values of 
$t^{\prime}$.

\end{description}

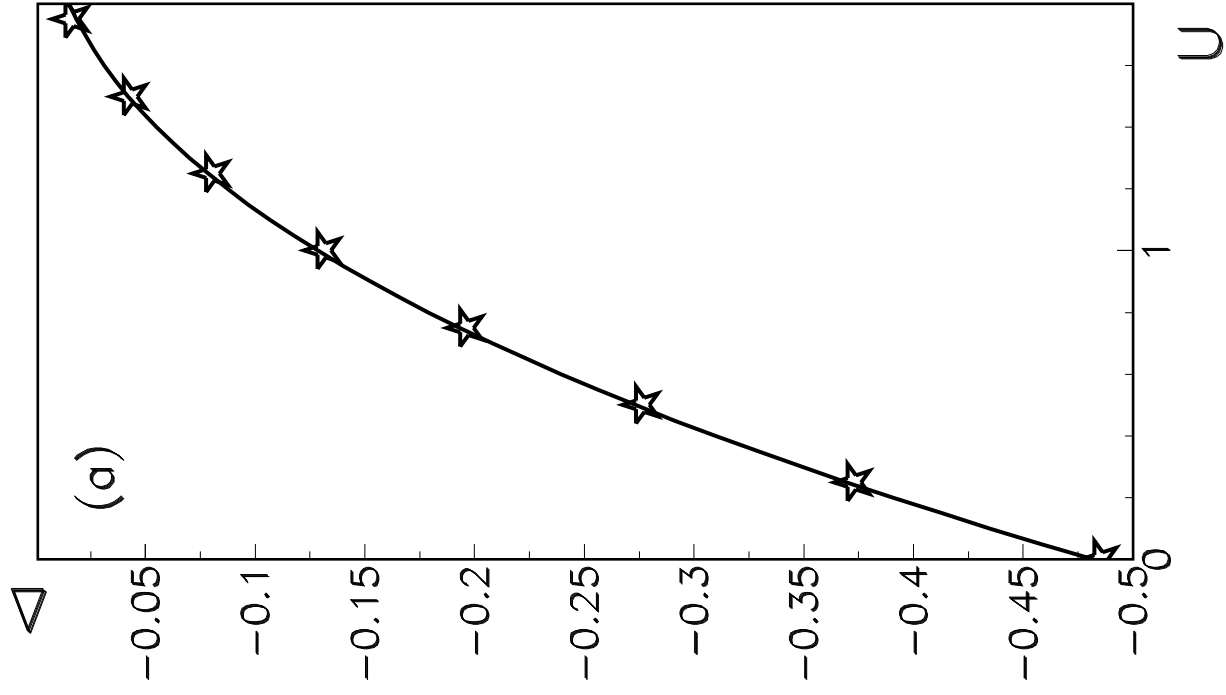
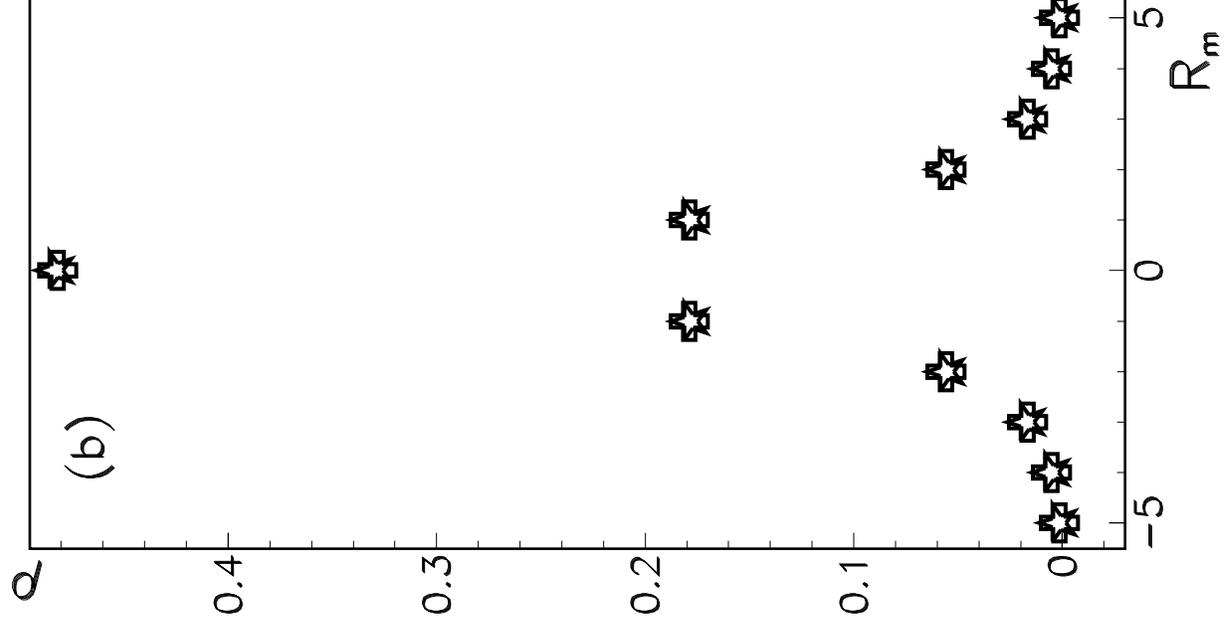
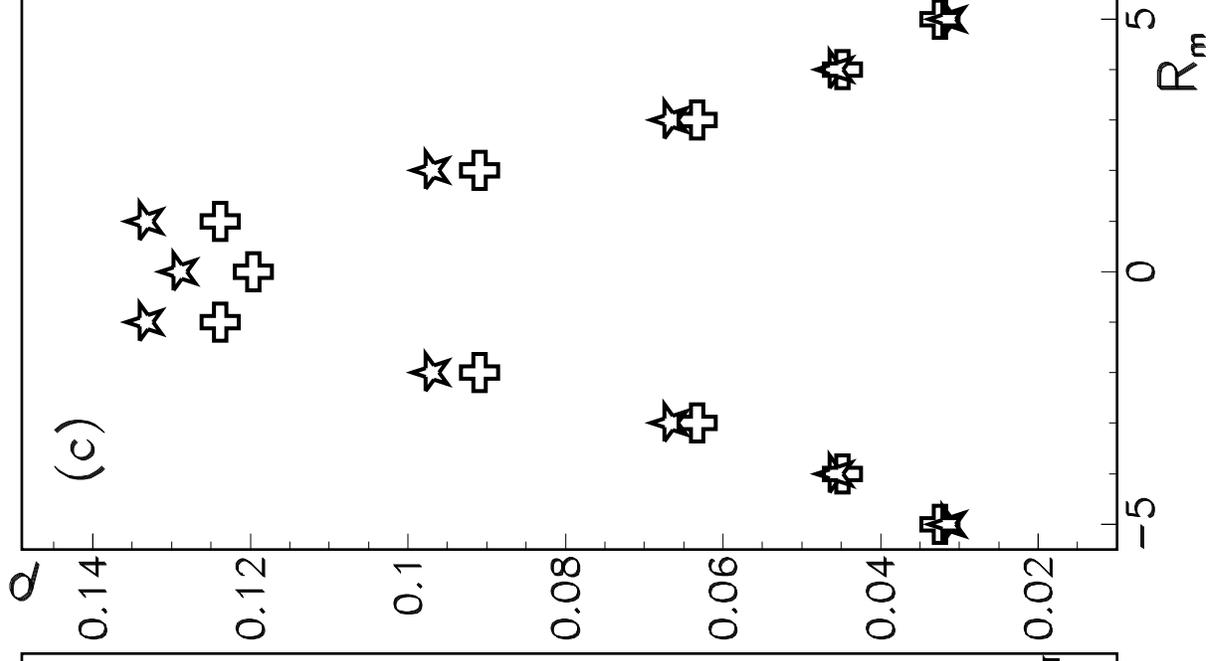
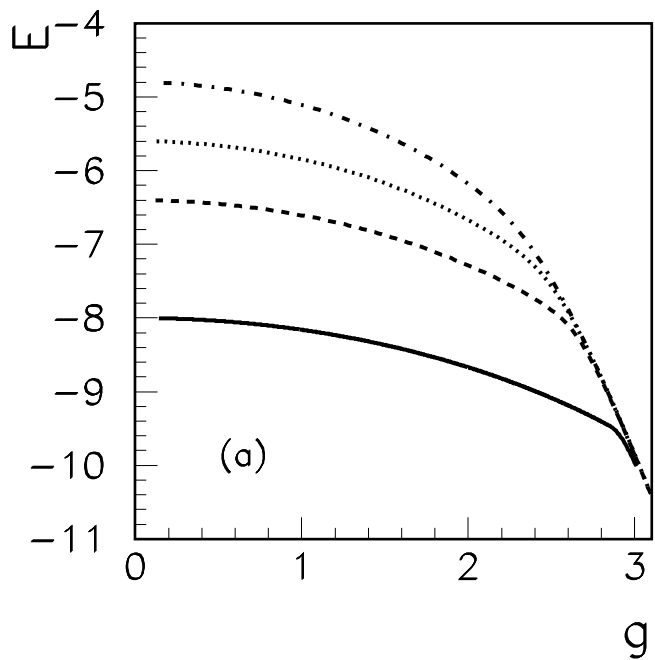
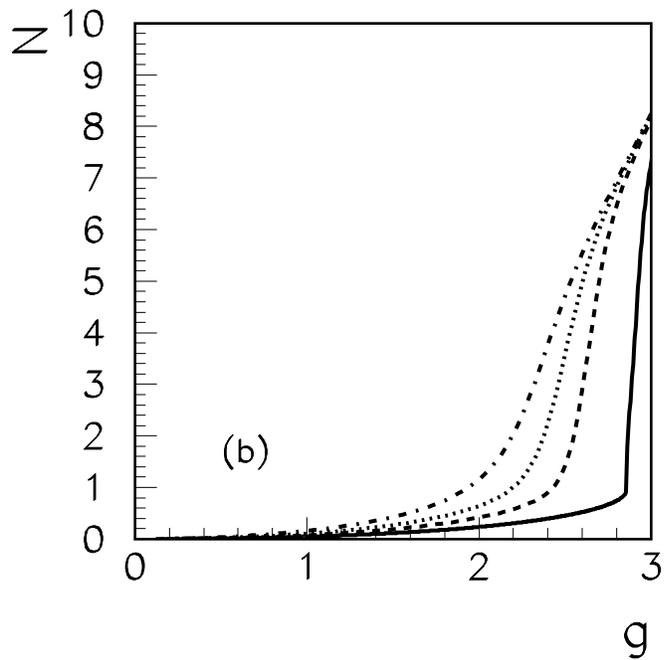
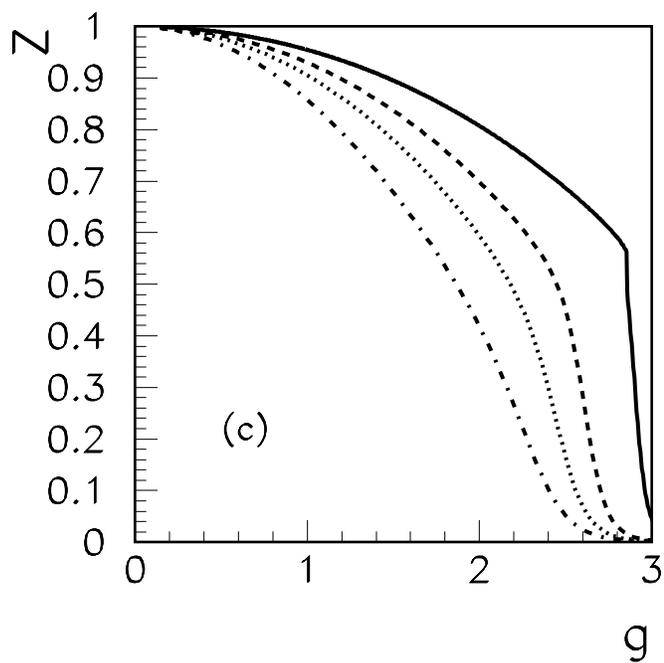
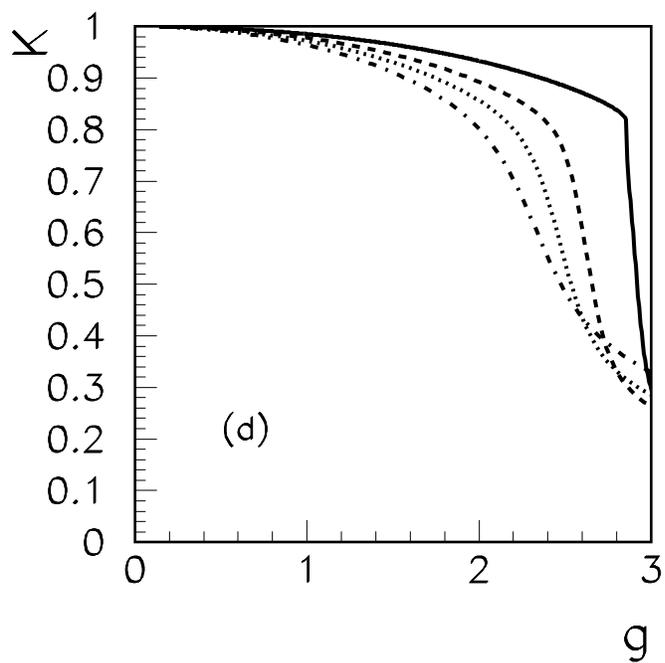
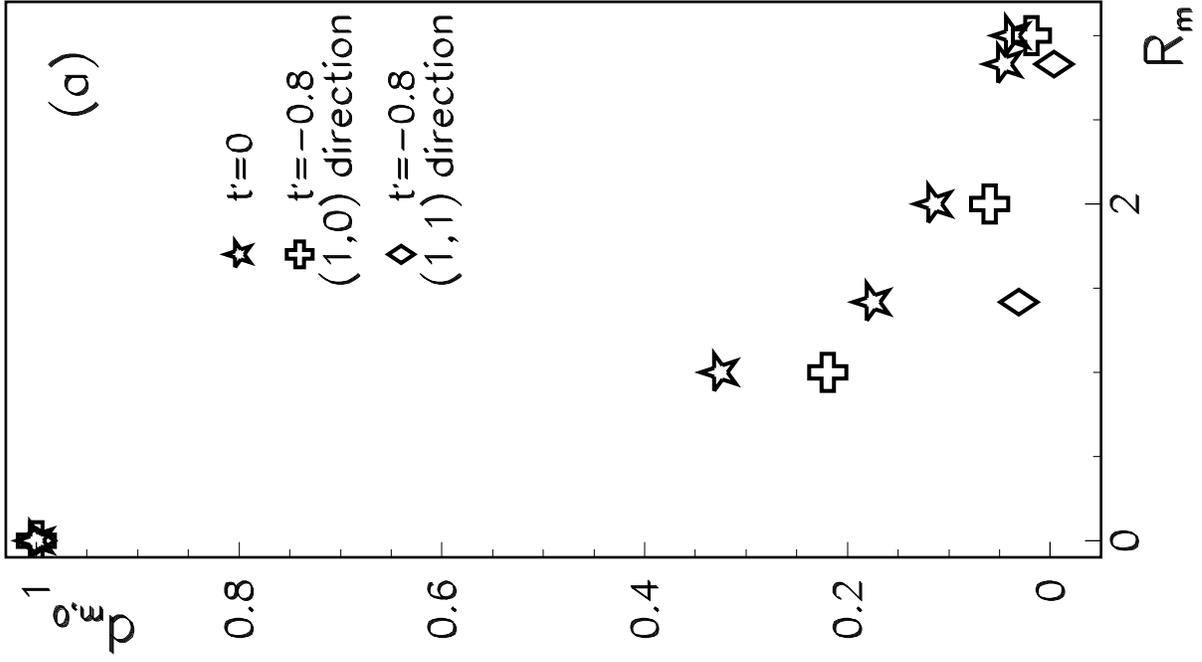
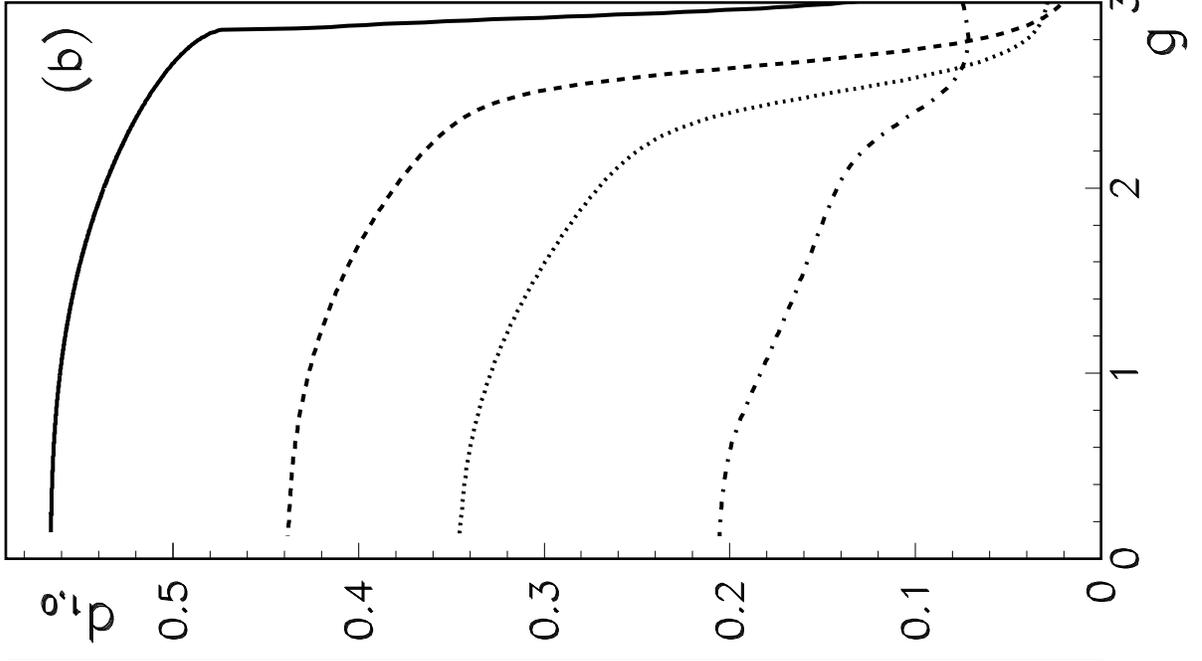
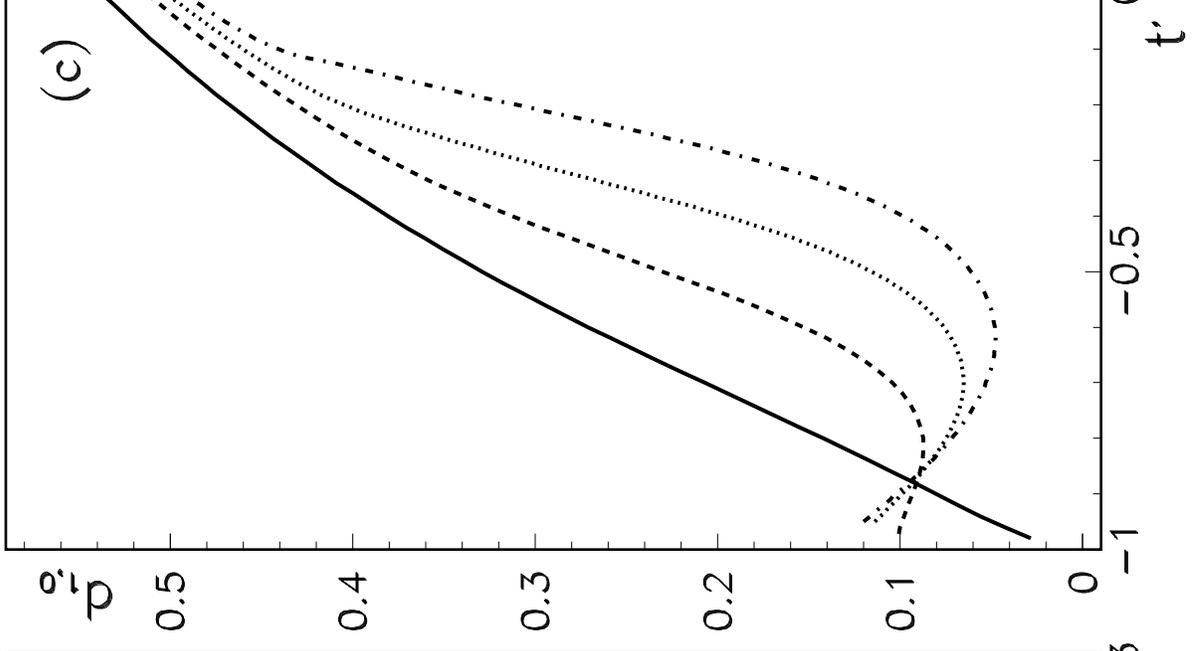
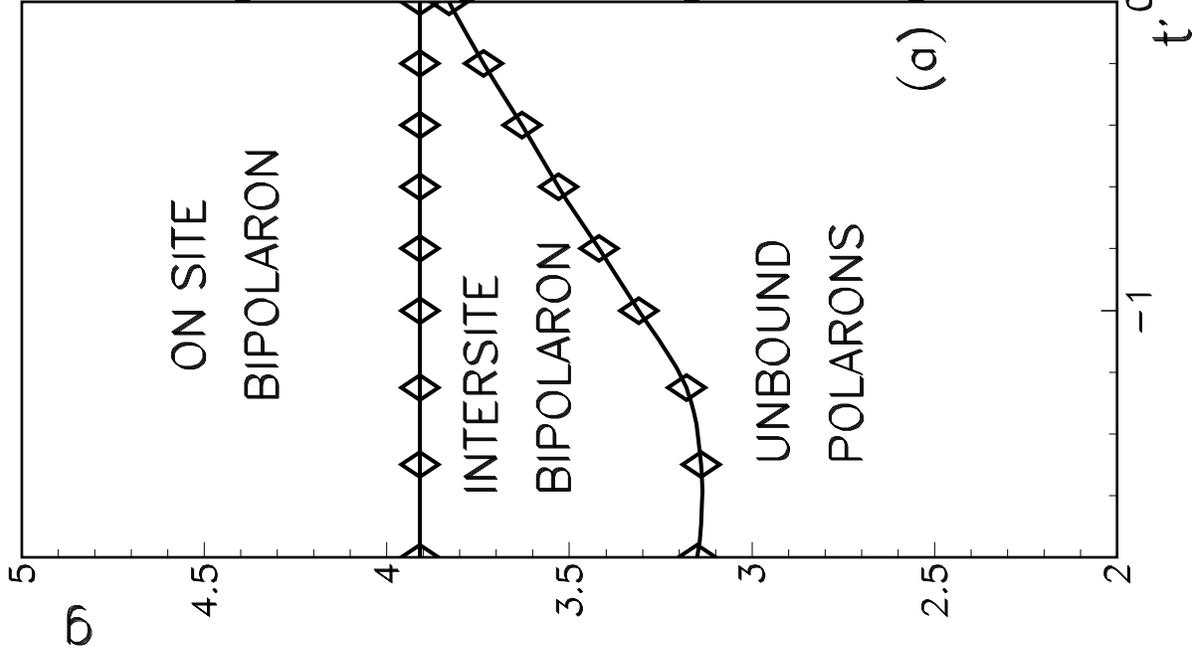
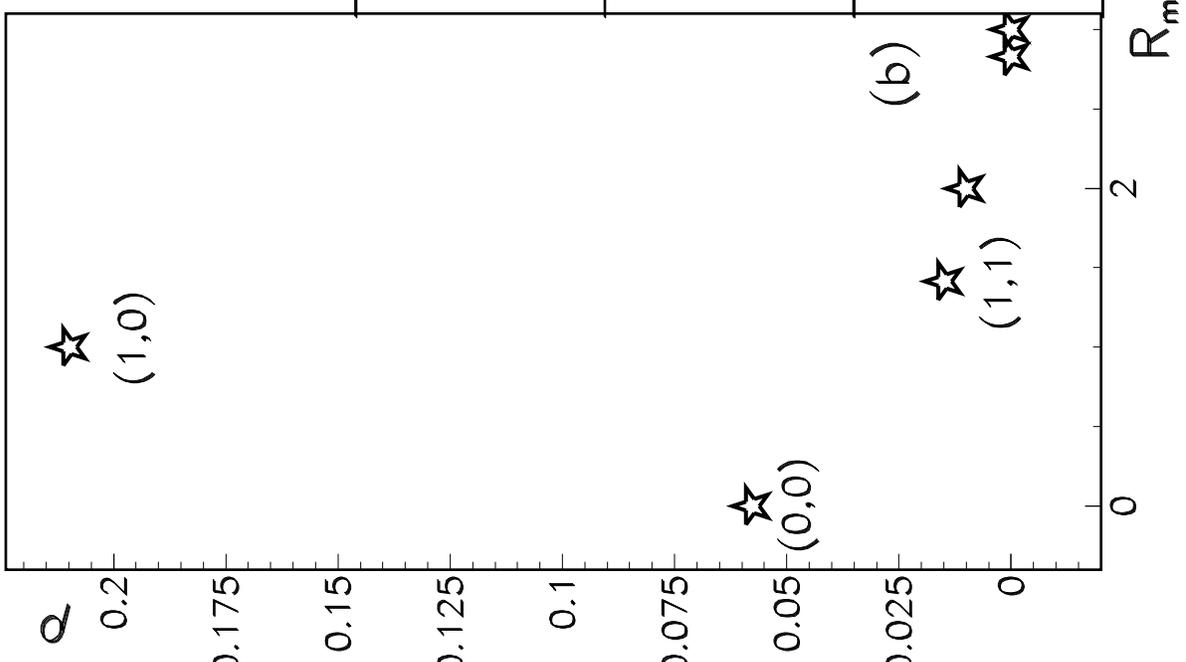
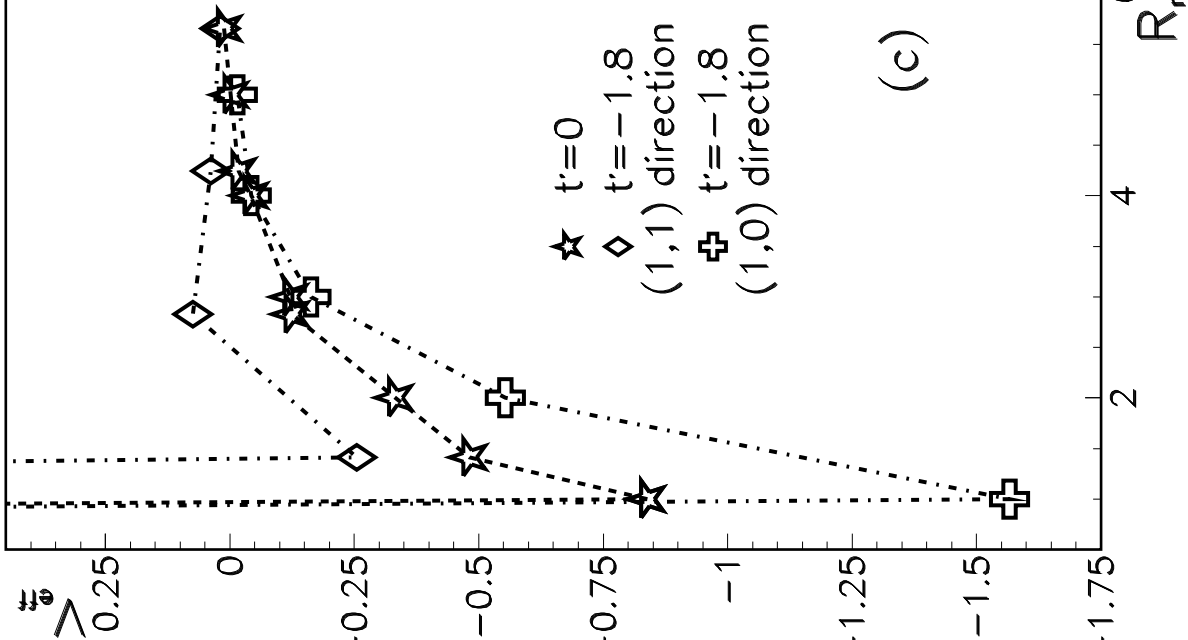
\end{document}